\documentstyle[12pt,epsf]{article}
\begin{document}

\def\be{\begin{equation}}
\def\ee{\end{equation}}
\def\ba{\begin{eqnarray}}
\def\ea{\end{eqnarray}}

\title{Multiscaling of energy correlations
       in the random-bond Potts model}
\author{Jesper Lykke Jacobsen \\
        LPTMS, b\^{a}timent~100,
        Universit\'e Paris-Sud, F-91405 Orsay, France}
\date{December 1999}
\maketitle

\begin{abstract}
We numerically calculate the exponent for the disorder averaged and
fixed-sample decay of the energy-energy correlator in the $q$-state
random-bond Potts model.
Our results are in good agreement with a two-loop expansion around
$q=2$ recently found from perturbative renormalisation group techniques,
fulfill the correlation length bound $\nu \ge 2/d$, and give further
evidence against replica symmetry breaking in this class of models.
\end{abstract}

\section{Introduction}

The $q$-state random-bond Potts model is an interesting framework for
examining how a phase transition is modified by quenched disorder coupling
to the local energy density. For $q>2$ such randomness acts as a relevant
perturbation \cite{Harris}, and for $q>4$ it even changes the nature of
the transition from first to second order (see Ref.~\cite{Cardy99} for
a review). In the regime where $(q-2)$ is small, a score of analytical
results have been obtained from the perturbative renormalization group,
and the various expansions for the central charge and the multiscaling
exponents for the moments of the spin-spin correlator compare convincingly
to recent numerical work \cite{Berche99}.

A particularly useful way of carrying out these simulations is to
consider the finite-size scaling of the Lyapunov spectrum of the
(random) transfer matrix, thus generalizing the method commonly applied
to the eigenvalue spectrum in a pure system \cite{Cardy98}.
A definite advantage over the more
traditional technique of Monte Carlo simulations \cite{Picco98} is that the
transfer matrices allow for a representation in which $q$ can be regarded
as a continuously varying parameter \cite{Blote82,Cardy98}, and in particular
one can study small non-integer values of $(q-2)$.

The outcome of applying this method to the {\em energetic} sector of the
transfer matrix, however, led to contradictory results \cite{Cardy98}.
Most notably, the exponent $\tilde{X}_1$ describing the asymptotic decay
of the disorder-averaged first moment of the two-point function
$\overline{\langle \varepsilon(x_1) \varepsilon(x_2) \rangle} \propto
 |x_1 - x_2|^{-2 \tilde{X}_1}$
seemed to be a rapidly decreasing function of $q$, in sharp disagreement
with an exact bound on the correlation length exponent,
$\nu \ge 2/d$ \cite{Chayes}, which in our notation reads $\tilde{X}_1 \ge 1$.

More recent numerical work has emphasized the importance of crossover
behavior \cite{Picco98} from the random fixed point to, on one side,
the pure Potts model and, on the other, a percolation-like limit
\cite{Cardy98} in which the ratio $R=K_1/K_2$ between strong and weak
couplings tends to infinity. It became clear that while the fixed ratio
$R=2$ employed in Ref.~\cite{Cardy98} seems to have been adequate for
studying the spin sector when $(q-2)$ is small, in general higher values
of $R$ are needed to measure the true random behavior in the regime $q>4$
\cite{Picco98}.

These findings were put on a firmer ground when it was realized
\cite{Picco99} that Zamolodchikov's $c$-theorem \cite{Zamolodchikov} can
be used to explicitly trace out the critical disorder strength $R_*(q)$ as a
function of $q$, by scanning for an extremum of the effective central
charge. In conjunction with an improved transfer matrix algorithm in
which the Potts model is treated through its representation as a loop
model, this allowed the authors of Ref.~\cite{Picco99} to produce very
accurate results for the central charge and the magnetic scaling dimension
in the regime $q\gg4$.

On the analytical side, the perturbative expansions for the first three
moments of the energetic two-point function have been known for quite
some time \cite{Ludwig90}. It was however only very recently that 
Jeng and Ludwig \cite{Ludwig99} succeded in generalizing these
computations to a general $N$th moment of the energy operator
$\overline{\langle \varepsilon(x_1) \varepsilon(x_2) \rangle^N} \propto
 |x_1 - x_2|^{-2 \tilde{X}_N}$, yielding
\be
 \tilde{X}_N = N \left(1 - \frac{2}{9 \pi^2} (3N - 4)(q-2)^2 +
                 {\cal O}(q-2)^3 \right).
 \label{two-loop}
\ee
In particular this makes available the experimentally relevant exponent
$\tilde{X}'_0$ describing the typical decay of the energy-energy correlator
in a fixed sample at criticality \cite{Ludwig90}.

In the present publication we show that by combining the methods of
Refs.~\cite{Cardy98,Picco99}
the exponents $\tilde{X}_1$ and $\tilde{X}'_0$
can be quite accurately determined numerically for small
$(q-2)$. In particular we find $\tilde{X}_1 \ge 1$ in full agreement with
the correlation length bound \cite{Chayes}, and our results
lend strong support to the above two-loop results of
the perturbative renormalisation group.

\section{The simulations}

In order to compare our results with those of the $(q-2)$-expansion, while
on the other hand staying comfortably away from $q=2$ where logarithmic
corrections are expected \cite{Cardy86b}, our main series of data has
$q=2.5$. Iterating the transfer matrix for a strip of width $L$ a large
number $M \gg L$ of times, we examine the probability distribution of the
ratio between the two largest Lyapunov exponents $\Lambda_0$, $\Lambda_1$
\cite{Furstenberg} in terms of the free energy
gap $\Delta f(L) = \frac{1}{L M} \ln(\Lambda_0/\Lambda_1)$. We employ the loop
representation of the transfer matrix where each loop on the surrounding
lattice is given a weight $n = \sqrt{q}$ \cite{Baxter82}, and bond randomness
is incorporated by weighing the two possible vertex configurations by
$w_i$ and $1/w_i$, where $w_i$ is a quenched random variable that can assume
two different values $s$ and $1/s$, each one with probability $1/2$
\cite{Picco99}. By construction, the system is then on average at its
self-dual point \cite{Kinzel}. The strength of the disorder is
measured by $s>1$, which is related to the ratio between strong and weak
bonds by $R=K_1/K_2=\ln(1+s\sqrt{q})/\ln(1+\sqrt{q}/s)$. The maximum strip
width employed in the study is $L_{\rm max}=12$.

Following Ref.~\cite{Picco99}, we start by locating the critical disorder
strenght $s_*$ by searching for a maximum of the effective central charge.
To do so, we must be able to determine finite-size estimates $c(L,L+2)$
\cite{Cardy86} with five significant digits, which means that the free
energy $f_0(L) = \frac{1}{L M} \ln(\Lambda_0)$ must be known with seven digit
precision. These considerations fix the necessary number of iterations
to be $M=10^8$.

\begin{table}
 \begin{center}
 \begin{tabular}{r|ccc}
 $s$ & $c(6,8)$ & $c(8,10)$ & $c(10,12)$ \\ \hline
 1.0 & 0.63775  & 0.64844   & 0.65404    \\
 1.5 & 0.63774  & 0.64854   &            \\
 2.0 & 0.63770  & 0.64874   &            \\
 2.1 & 0.63767  & 0.64877   &            \\
 2.2 & 0.63764  & 0.64879   & 0.65415    \\
 2.3 & 0.63759  & 0.64880   & 0.65415    \\
 2.4 & 0.63754  & 0.64879   & 0.65413    \\
 2.5 & 0.63748  & 0.64879   & 0.65411    \\
 2.6 & 0.63740  & 0.64877   & 0.65409    \\
 \end{tabular}
 \end{center}
 \protect\caption[2]{\label{tab:c}Effective central charge $c(L,L+2)$ of the
 $q=2.5$ state model, as a function of disorder strength $s$ and the strip
 width $L$.}
\end{table}

Our results for $c(L,L+2)$ as a function of $L$ and $s$ are shown
in Table~\ref{tab:c}. For a sufficiently large system size $L$, these data
exhibit a maximum as a function of $s$, the position of which determines
a finite-size estimate $s_*(L)$, which converges to $s_*$ as $L\to\infty$.
From the data of Table~\ref{tab:c}, supplemented by improved three-point
fits \cite{Cardy98} (not shown), we extrapolate to $s_*(q=2.5) = 2.5(1)$.

The fluctuations in $\Delta f(L)$ are examined by dividing the strip into
$M/m$ samples, each one of length $m=10^5$, from which the first few
cumulants of $\Delta f(L)$ can be determined. As discussed in
Ref.~\cite{Cardy98}, the exponent $\tilde{X}'_0$ is related to the
finite-size scaling \cite{Cardy83} of the mean value (first cumulant)
of $\Delta f(L)$, whereas $\tilde{X}_1$ is similarly determined from
the sum of the entire cumulant expansion. In practice, the second cumulant
is roughly two orders of magnetude smaller than the first, and higher
cumulants are
expected to be further suppressed, even though their determination is
made difficult by numerical instabilities. We can therefore with
confidence truncate the sum of the cumulants after the second one.

\begin{table}
 \begin{center}
 \begin{tabular}{r|cccc}
 $s$ & $\tilde{X}'_0(6)$  & $\tilde{X}'_0(8)$  &
       $\tilde{X}'_0(10)$ & $\tilde{X}'_0(12)$ \\ \hline
 1.0 & 0.9791 & 0.9513 & 0.9375 & 0.9293 \\
 1.5 & 1.0026 & 0.9754 & 0.9622 &        \\
 2.0 & 1.0468 & 1.0191 & 1.0057 &        \\
 2.2 & 1.0662 & 1.0377 & 1.0238 & 1.0158 \\
 2.3 & 1.0760 & 1.0470 & 1.0328 & 1.0246 \\
 2.4 & 1.0859 & 1.0563 & 1.0418 & 1.0333 \\
 2.5 & 1.0958 & 1.0656 & 1.0506 & 1.0419 \\
 2.6 & 1.1057 & 1.0749 & 1.0595 & 1.0504 \\
 \end{tabular}
 \end{center}
 \protect\caption[2]{\label{tab:x0}Effective exponent $\tilde{X}'_0(L)$ of the
 $q=2.5$ state model, as a function of disorder strength $s$ and the strip
 width $L$.}
\end{table}

\begin{table}
 \begin{center}
 \begin{tabular}{r|cccc}
 $s$ & $\tilde{X}_1(6)$  & $\tilde{X}_1(8)$  &
       $\tilde{X}_1(10)$ & $\tilde{X}_1(12)$ \\ \hline
 1.0 & 0.9791 & 0.9513 & 0.9375 & 0.9293 \\
 1.5 & 1.0016 & 0.9746 & 0.9614 &        \\
 2.0 & 1.0389 & 1.0124 & 0.9997 &        \\
 2.2 & 1.0534 & 1.0269 & 1.0142 & 1.0065 \\
 2.3 & 1.0603 & 1.0338 & 1.0212 & 1.0133 \\
 2.4 & 1.0671 & 1.0405 & 1.0279 & 1.0198 \\
 2.5 & 1.0735 & 1.0469 & 1.0343 & 1.0260 \\
 2.6 & 1.0798 & 1.0531 & 1.0405 & 1.0320 \\
 \end{tabular}
 \end{center}
 \protect\caption[2]{\label{tab:x1}Effective exponent $\tilde{X}_1(L)$ of the
 $q=2.5$ state model, as a function of disorder strength $s$ and the strip
 width $L$.}
\end{table}

Resulting finite-size estimates \cite{Cardy83} of $\tilde{X}'_0$ and
$\tilde{X}_1$ are shown in Tables~\ref{tab:x0} and \ref{tab:x1} respectively.
Unlike what seemed to be the situation in the {\em magnetic} sector, these
estimates exhibit a pronounced dependence on $s$. Ref.~\cite{Cardy98}
worked at fixed $R=2$, which for $q=2.5$ would correspond to $s\simeq 1.7$,
and found $\tilde{X}'_0 < 1$ for all $q>2$. We see here that the correct
way to extract these exponents is to extrapolate the $s=s_*$ data to the
$L\to\infty$ limit. With the help of improved two-point estimates
\cite{Cardy98} (not shown) we thus obtain
\be
  \tilde{X}'_0 = 1.02(1) \ \ \ \
  \tilde{X}_1 = 1.00(1),
\ee
which verifies the bound of Ref.~\cite{Chayes}.
These exponents, as well as the result for their difference
$\tilde{X}'_0 - \tilde{X}_1 = 0.015(5)$, are in very good agreement with
the $(q-2)$-expansion; see Eq.~(\ref{two-loop}).

We have also performed simulations for higher values of $q$, where the
discrepancy between Refs.~\cite{Cardy98} and \cite{Chayes} was even more
pronounced, since $s_*$ is an increasing function of $q$.
For $q=2.75$ and $q=3$, we had to increase the number of iterations
to $M=10^9$ in order to keep the error bars under control despite the
higher disorder strength. In all cases we found good agreement with
Ref.~\cite{Chayes} and with the $(q-2)$-expansion, at least in the
range where the latter can be assumed to be valid. A summary of our
results is given in Table~\ref{tab:res}.

\begin{table}
 \begin{center}
 \begin{tabular}{r|ll|ll|ll}
 $q$  & $s_*$  & $R_*$   & $\tilde{X}'_0$ &  & $\tilde{X}_1$ &   \\
      &        &         & numerics & theory & numerics & theory \\ \hline
 2.50 & 2.5(1) & 3.3(2)  & 1.02(1) & 1.023 & 1.00(1) & 1.006 \\
 2.75 & 3.0(3) & 4.1(5)  & 1.04(2) & 1.051 & 1.01(1) & 1.013 \\
 3.00 & 3.5(5) & 4.7(10) & 1.06(3) & 1.09  & 1.02(2) & 1.02  \\
 \end{tabular}
 \end{center}
 \protect\caption[2]{\label{tab:res}Numerical results for the critical
 disorder strength ($s_*$ or $R_*$) and energetic scaling exponents
 ($\tilde{X}_0'$ and $\tilde{X}_1$) as functions of $q$. The agreement
 with the two-loop expansion, Eq.~(\ref{two-loop}), is good.}
\end{table}

\section{Conclusion}

In summary, we have shown that the apparent violation of the correlation
length bound \cite{Chayes} observed in Ref.~\cite{Cardy98} can be dismissed
as a crossover effect due to the lack of tuning to the critical disorder
strength. In conjunction with the results on degeneracy and descendents
given in Ref.~\cite{Cardy98} we would thus claim that the transfer matrix
method \cite{Cardy98,Picco99} can, at least in principle, be used to relate
the entire Lyapunov spectrum to the operator content of the (as yet
unknown) underlying conformal field theory.

In particular, we have supplied convincing numerical validation of the
two-loop expansion (\ref{two-loop}) for the energetic multiscaling
exponents \cite{Ludwig99}. Our results also provide further evidence in
favour of the replica symmetric approach to the perturbative calculations,
since the assumption of initial replica symmetry breaking leads to
$\tilde{X}_1 = 1 + {\cal O}(q-2)^3$ \cite{DDPP95}, which seems to be
ruled out by the results given in Table~\ref{tab:res}.

\end{document}